\documentclass[onecollarge,final]{svjour2}
\usepackage[numbers, sort&compress]{natbib}

\smartqed  
\usepackage{graphicx}
\usepackage{color}
\usepackage{amsmath,amssymb,bm}
\usepackage{amsfonts}
\usepackage{dcolumn}
\usepackage{ulem}
\graphicspath{{figs/}{.}}

\newcommand{\nuc}[2]{^{#1}\mathrm{#2}}
\newcommand{\Lag}[0]{\ensuremath{\mathcal{L}}}
\newcommand{\mrm}[1]{\ensuremath{\mathrm{#1}}}

\newcommand{\mbf}[1]{\ensuremath{\mathbf{#1}}}
\newcommand{\klo}[0]{\ensuremath{k_\mrm{lo}}}
\newcommand{\khi}[0]{\ensuremath{k_\mrm{hi}}}
\renewcommand{\d}[0]{\ensuremath{\mathrm{d}}}
\begin{document}
\title{Three-body halo states in effective field theory
}


\author{Emil Ryberg \and
  Christian Forss\'en  \and
  Lucas Platter
}


\institute{E. Ryberg \and C. Forss\'en \at
              Department of  Physics, Chalmers University
              of Technology, SE-412 96 G\"oteborg, Sweden
              \and
              L. Platter \at
              Department of Physics and Astronomy, University of
              Tennessee, Knoxville, TN 37996, USA\\
              Physics Division, Oak Ridge National Laboratory, Oak Ridge,
              TN 37831, USA  
}

\date{Received: date / Accepted: date}

\maketitle

\begin{abstract}
  In this paper we study the renormalization of Halo effective field
  theory applied to the Helium-6 halo nucleus seen as an
  $\alpha$-neutron-neutron three-body state. We include the $0^+$
  dineutron channel together with both the $3/2^-$ and $1/2^-$
  neutron-$\alpha$ channels into the field theory and study all of the
  six lowest-order three-body interactions that are
  present. Furthermore, we discuss three different prescriptions to
  handle the unphysical poles in the P-wave two-body sector. In the
  simpler field theory without the $1/2^-$ channel present we find
  that the bound-state spectrum of the field theory is renormalized by
  the inclusion of a single three-body interaction. However, in the
  field theory with both the $3/2^-$ and $1/2^-$ included, the system
  can not be renormalized by only one three-body operator.
\end{abstract}

\section{Introduction}
\label{intro}
%
Effective field theory (EFT) has become a widely used tool to
construct interactions for few-body problems in atomic and nuclear
physics. Examples are the nucleon-nucleon interaction constructed
using Chiral EFT \cite{Epelbaum:2008ga,Machleidt:2011zz}
and the atom-atom interaction for systems with a large scattering
length \cite{Braaten:2004rn,Hammer:2012id}. The properties of the
corresponding three-body systems are usually obtained by the solution
of the Faddeev equation. Halo nuclei
\cite{Tanihata:1986kh,2004PhR...389....1J} have become another arena
for the application of EFTs. These are systems of tightly bound cores
with weakly bound valence nucleons that can be found close to the
neutron and proton driplines. Halo EFT uses the ratio of the
valence-nucleon separation energy and the binding of the core as the
expansion parameter for the low-energy EFT expansion.  The core and
valence nucleons are then the effective degrees of freedom used within
this approach and the complexity of the problem is significantly
reduced. An advantage of this approach is that the uncertainties of
the model can be systematically reduced, by including more terms in
the low-energy expansion.

Halo EFT has previously been applied to resonant two-body systems
\cite{Bertulani:2002sz,Bedaque:2003wa,Higa:2008dn,Brown:2013zla},
one-neutron halos
\cite{Hammer:2011ye,Rupak:2012cr,Acharya:2013nia,Zhang:2013kja} and
one-proton halos \cite{paper:A,Zhang:2014zsa,paper:B,paper:C}.  The
ground state of Helium-6 has also been analyzed in Halo EFT, both by
Rotureau and van Kolck~\cite{Rotureau:2012yu} and by Ji, Elster and
Phillips~\cite{Ji:2014wta}. Additional efforts on two-neutron halos
can be found in Refs.~\cite{Acharya:2013aea,Hagen:2013xga,Hagen:2013jqa}. 

One major tenet of EFTs is that observables have to be independent of
any short-distance regulators. Such regulators have to be introduced
to deal with the singularities that a low-momentum expansion typically
introduces. One important consequence of this requirement is that
operators frequently have to be promoted to lower order than naively
expected from their mass dimension. One well-known example is the
three-boson system where naive dimensional analysis implies that the
three-body interaction should enter at N$^2$LO. However, as was shown
in an EFT treatment in Ref.~\cite{Bedaque:1998km} this interaction is
needed already at LO for the theory to be renormalized.

Another problem that can occur in few-body systems is the appearance
of spurious bound states in the two-body subsystem at cutoffs larger
than the breakdown scale. This happens for example in the deuteron
system when a chiral potential with a large momentum-space cutoff is
employed, but also in Halo EFT when systems that interact resonantly
in a relative P-wave are considered. For example, in the case of a
low-lying P-wave two-body resonance one can fix the P-wave scattering
length and effective range such that the two-body propagator
reproduces this resonance. However, the denominator of such a
propagator is then an order-three polynomial, which implies that
additional two-body poles are present in the theory.

The Halo EFT analysis of the Helium-6 system is one example for which
spurious bound states have become an immediate problem. The
neutron-$\alpha$ interaction is resonant in the P-wave and the
resulting two-body t-matrix has three poles in the complex plane with
one of them being unphysical. In previous efforts it was suggested to
remove the unitarity piece $ik^3$ from the denominator of the P-wave
propagator, see Ref.~\cite{Bedaque:2003wa}. This prescription is
possible if only one fine-tuning is assumed and it is restricted to
LO. We propose and discuss two additional prescriptions.

In this paper, we derive (for the first time) the complete
field-theoretical equations for the Helium-6 system from appropriate
diagrammatics. The two-body sector is renormalized by fitting the
low-energy constants to the resonance position and width of the
Helium-5 system and the neutron-neutron scattering length. In
Refs.~\cite{Rotureau:2012yu,Ji:2014wta}, the Helium-6 system was
renormalized with the simplest possible three-body counterterm and an
analysis of other possible operators was omitted. Here, we perform a
renormalization analysis of the bound-state spectrum, using all of the
six lowest-order three-body interactions that are present. We also
compare the three different prescriptions of how to treat the
unphysical P-wave poles in the two-body sector.

This manuscript is organized as follows: In Sec.~\ref{sec:method}, we
discuss the EFT used to describe the Helium-6 system and its
renormalization in the two-body sector. In Sec.~\ref{sec:results} we
analyze the renormalization of the three-body system. Finally, a
conclusion is provided in Sec.~\ref{sec:conclusion}.

\section{Method}
\label{sec:method}

In this section, we present a framework for the treatment of the bound
$0^+$ ground state of $\nuc{6}{He}$, which has a two-neutron
separation energy of $1~\mrm{MeV}$. It can be viewed as consisting of
an $\alpha$ particle and two neutrons. The one-nucleon separation
energy of the $\alpha$ particle is $20~\mrm{MeV}$ and this defines the
break-down scale in energy. Thus we have a good separation of
scales. We include the S-wave dineutron channel and both the $3/2^-$
and $1/2^-$ channels of the P-wave neutron-$\alpha$ interaction. These
two channels correspond to the two low-lying resonances of
$\nuc{5}{He}$, with energy positions and widths
$E^{(3/2^-)}=0.798~\mrm{MeV}$, $\Gamma^{(3/2^-)}=0.648~\mrm{MeV}$,
$E^{(1/2^-)}=2.07~\mrm{MeV}$ and $\Gamma^{(1/2^-)}=5.57~\mrm{MeV}$,
respectively~\cite{Tilley:2002vg}.  We expect that the $3/2^-$ is more important than the
$1/2^-$ channel, since it is lower in energy.

\subsection{Lagrangian}
The fields that are included in this field theory are the $1/2^+$
neutron, $n_\sigma$, the $0^+$ $\alpha$ core, $c$, the $0^+$ dineutron
field, $b$, the $\nuc{5}{He}(3/2^-)$ field, $d_a$, and the
$\nuc{5}{He}(1/2^-)$ field, $\tilde{d}_\sigma$. The spin indices are
defined as $\sigma=-1/2,1/2$ and $a=-3/2,-1/2,1/2,3/2$. We will also
use $\chi$ as a spin-$1/2$ and $b$ as a spin-$3/2$ index, together
with $i,j=-1,0,1$ as spin-$1$ indices.

We write the Lagrangian for the $0^+$ channel of $\nuc{6}{He}$ as a
sum of Lagrangian parts
\begin{equation}
\Lag=\Lag^{(1)}+\Lag^{(2)}+\Lag^{(3)}~.
\label{eq:Lag3body123}
\end{equation}
The one-body part is given by
\begin{equation}
\Lag^{(1)}=c^\dagger\left[i\partial_t+\frac{\nabla^2}{8m}\right]c
+n_\sigma^\dagger\left[i\partial_t+\frac{\nabla^2}{2m}\right]n_\sigma+\dots~,
\label{eq:Lag3body1}
\end{equation}
where the neutron, $n_\sigma$, has a mass $m$ and the core, $c$, has a
mass $4m$. 
The dots refer to relativistic one-body corrections.

The two-body Lagrangian is given by
\begin{align}
\Lag^{(2)}=&d^\dagger_a\left[\Delta_1+\nu_1\left(i\partial_t+\frac{\nabla^2}{10m}\right)\right]d_a+g_1\mathcal{C}^a_{i\chi}\left[d^\dagger_ac\left(\frac{4}{5}i\overrightarrow{\nabla}_i-\frac{1}{5}i\overleftarrow{\nabla}_i\right)n_\chi+\mrm{h.c.}\right]
\nonumber\\
&+\tilde{d}^\dagger_\sigma\left[\tilde{\Delta}_1+\tilde{\nu}_1\left(i\partial_t+\frac{\nabla^2}{10m}\right)\right]\tilde{d}_\sigma+\tilde{g}_1\mathcal{C}^\sigma_{i\chi}\left[\tilde{d}^\dagger_\sigma c\left(\frac{4}{5}i\overrightarrow{\nabla}_i-\frac{1}{5}i\overleftarrow{\nabla}_i\right)n_\chi+\mrm{h.c.}\right]
\nonumber\\
&+b^\dagger\Delta_0b+\frac{1}{2}g_0\mathcal{C}^0_{\sigma\chi}\left(b^\dagger n_\sigma n_\chi+\mrm{h.c.}\right)+\dots~.
\label{eq:Lag3body2}
\end{align}
The parameters $\Delta_1$, $\nu_1$ and $g_1$ define the
neutron-$\alpha$ P-wave interaction in the $3/2^-$ channel and
$\tilde{\Delta}_1$, $\tilde{\nu}_1$ and $\tilde{g}_1$ define the
neutron-$\alpha$ P-wave interaction in the $1/2^-$ channel. The
low-energy constants $\Delta_0$ and $g_0$ define the neutron-neutron
S-wave interaction in the $0^+$ channel. We define the Clebsch-Gordan
coefficients accoring to
$\mathcal{C}^{a}_{i\chi}=\langle1i \frac{1}{2}\chi
|(1\frac{1}{2})\frac{3}{2}a \rangle$,
$\mathcal{C}^\sigma_{i\chi}=\langle 1i \frac{1}{2}\chi
|(1\frac{1}{2})\frac{1}{2}\sigma \rangle$ and
$\mathcal{C}^{0}_{\sigma\chi}=\langle\frac{1}{2}\sigma\frac{1}{2}\chi
|(\frac{1}{2}\frac{1}{2})00 \rangle$. The derivative operators
$(\frac{4}{5}i\overrightarrow{\nabla}_i-\frac{1}{5}i\overleftarrow{\nabla}_i)$
ensure that the interactions are in the P-wave channel and that the
Lagrangian obeys Galilean invariance. The dots refer to higher-order
two-body interactions.

For the three-body part of the Lagrangian we have
\begin{align}
\Lag^{(3)}=&h_1 \mathcal{C}^0_{a\chi i}\mathcal{C}^0_{a'\chi' i'}\left(d_a^\dagger i\overleftrightarrow{\nabla}_in_\chi^\dagger\right) \left(d_{a'} i\overleftrightarrow{\nabla}_{i'}n_{\chi'}\right)
\nonumber\\
&+h_2 \mathcal{C}^0_{\sigma\chi i}\mathcal{C}^0_{a'\chi' i'}\left[\left(\tilde{d}_\sigma^\dagger i\overleftrightarrow{\nabla}_in_\chi^\dagger\right) \left(d_{a'} i\overleftrightarrow{\nabla}_{i'}n_{\chi'}\right)+\mrm{h.c.}\right]
\nonumber\\
&+h_3 \mathcal{C}^0_{\sigma\chi i}\mathcal{C}^0_{\sigma'\chi' i'}\left(\tilde{d}_\sigma^\dagger i\overleftrightarrow{\nabla}_in_\chi^\dagger\right) \left(\tilde{d}_{\sigma'} i\overleftrightarrow{\nabla}_{i'}n_{\chi'}\right)
\nonumber\\
&+h_4 b^\dagger c^\dagger b c
\nonumber\\
&+h_5\mathcal{C}^0_{a\chi i}\left[\left(d_a^\dagger i\overleftrightarrow{\nabla}_in_\chi^\dagger\right) bc+\mrm{h.c.}\right]
\nonumber\\
&+h_6 \mathcal{C}^0_{\sigma\chi i}\left[\left(\tilde{d}_\sigma^\dagger i\overleftrightarrow{\nabla}_in_\chi^\dagger\right) bc+\mrm{h.c.}\right]+\dots~.
\label{eq:Lag3body30}
\end{align}
Here we have defined the Galilean invariant P-wave interaction
operators as
$\overleftrightarrow{\nabla}=
\frac{5}{6}\overrightarrow{\nabla}-\frac{1}{6}\overleftarrow{\nabla}$. The
triple-subscript $\mathcal{C}$ objects are defined according to
$\mathcal{C}^0_{a\chi i}=\mathcal{C}_{a\chi}^{j}\mathcal{C}_{ji}^0$
and
$\mathcal{C}^0_{\sigma\chi
  i}=\mathcal{C}_{\sigma\chi}^{j}\mathcal{C}_{ji}^0$. The terms that
we have included in this three-body Lagrangian part
(\ref{eq:Lag3body30}) are the lowest order interactions that are of
scaling dimension $7$ (note that the p-wave dimer field has scaling
dimension 1 and the dineutron field has scaling dimension 2). As such,
they are expected to enter at N$^2$LO. However, in order to achieve
proper renormalization we need to promote at least one of them to
LO. We will discuss this further in Sec.~\ref{sec:results}. Note that
in previous treatments of $\nuc{6}{He}$ only the $h_1$ term has been
considered~\cite{Rotureau:2012yu,Ji:2014wta}.

\subsection{Two-body physics}

We begin by considering the two-body sector of the field theory. This
amounts to writing down the relevant dicluster propagators, that is
the dineutron propagator and the $\nuc{5}{He}$ propagators for the
$3/2^-$ and $1/2^-$ channels. We then discuss how to remove the
unphysical pole of the P-wave propagators, using one of three
prescriptions.

\begin{figure}[t!]
\begin{centering}
\includegraphics*[scale=0.7,angle=0,clip=true]{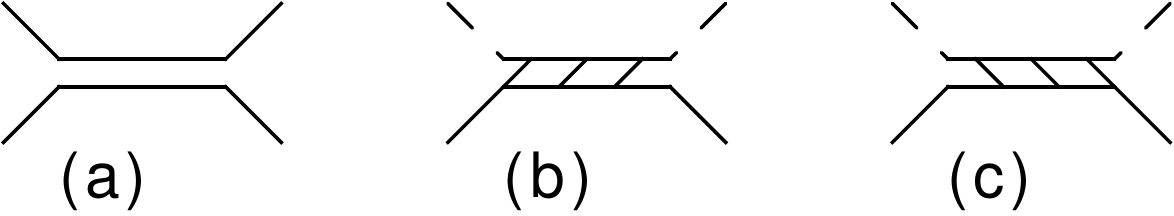}
\caption{Elastic scattering in the two-body subsystems of
  $\nuc{6}{He}$. (a) The neutron-neutron scattering diagram. The full
  dineutron propagator is denoted by the double line. (b) The
  neutron-$\alpha$ elastic scattering diagram in the $3/2^-$ channel. The
  full $\nuc{5}{He}(3/2^-)$ dicluster propagator is denoted by the
  double line with internal bottom-to-top right tilted lines. (c) The
  neutron-$\alpha$ elastic scattering diagram in the $1/2^-$ channel. The
  full $\nuc{5}{He}(1/2^-)$ dicluster propagator is denoted by the
  double line with internal bottom-to-top left tilted lines.}
\label{fig:twobodyprops}
\end{centering}
\end{figure}

\paragraph{The dineutron subsystem} -- For the neutron-neutron interaction we restrict ourselves to the
S-wave interaction. This interaction has an unnaturally large
scattering length, $a_0=-18.9~\mathrm{fm}$~\cite{Gardestig:2009ya}, which means that we must
resum the interaction to infinite order. The resulting full LO
dineutron propagator is therefore given by a geometric series, which
we write in closed form as
\begin{equation}
iB(E,\mbf{p})=\frac{i}{\Delta_0+\Sigma_0(E,\mbf{p})}~.
\label{eq:fulldineutronprop1}
\end{equation}
The irreducible self-energy, $\Sigma_0$, is given by
\begin{equation}
\Sigma_0(E,\mbf{p})=\frac{g_0^2m}{2\pi^2}\left[L_1-\frac{i\pi}{2}\sqrt{mE-\frac{\mbf{p}^2}{4}}\right]~,
\label{eq:irredselfenergydineutron}
\end{equation}
with a divergence $L_1$, defined by
\begin{equation}
L_n=\int\d p p^{n-1}~.
\label{eq:divint}
\end{equation}
This divergence is absorbed in the renormalization of the paramter
$\Delta_0$. The scattering diagram
is given  in Fig.~\ref{fig:twobodyprops}(a) where the double line defines the full dineutron
propagator. The propagator (\ref{eq:fulldineutronprop1}) is given for
a four-momentum $(E,\mbf{p})$.
%

By matching to ERE parameters, i.e. the neutron-neutron scattering
length $a_0$, we write the full dineutron propagator as
\begin{equation}
B(E,\mbf{p})=\frac{4\pi}{g_0^2m}\frac{1}{\frac{1}{a_0}+i\sqrt{m\left(E-\frac{\mbf{p}^2}{4m}\right)}},
\label{eq:fulldineutronprop2}
\end{equation}
which is the expression that we will use in actual calculations.

\paragraph{The $\nuc{5}{He}$ subsystem} -- For a P-wave interaction we need to also include the $\nu_1$ term of the Lagrangian part (\ref{eq:Lag3body2}) to achieve proper
renormalization. This term gives the effective range contribution. The LO full dicluster propagators for the P-wave
$3/2^-$ channel is thus written as
\begin{equation}
iD(E,\mbf{p})=\frac{i}{\Delta_1+\nu_1\left(E-\frac{\mbf{p}^2}{10m}\right)+\Sigma_1(E,\mbf{p})}~,
\label{eq:fulldiclusterprop3half1}
\end{equation}
where the irreducible self energy is given by
\begin{equation}
\Sigma(E,\mbf{p})=\frac{4g_1^2m}{15\pi^2}\left[L_3+\left(\frac{8mE}{5}-\frac{4\mbf{p}^2}{25}\right)L_1-\frac{i\pi}{2}\left(\frac{8mE}{5}-\frac{4\mbf{p}^2}{25}\right)^{3/2}\right]~.
\label{eq:irredselfenergy5He3half}
\end{equation}
Note that the propagator (\ref{eq:fulldiclusterprop3half1}) is written as a geometric series in closed form, similar to the dineutron propagator (\ref{eq:fulldineutronprop1}).
In the P-wave irreducible self-energy
(\ref{eq:irredselfenergy5He3half}) there are two independent
divergences, given by the divergent integrals $L_1$ and $L_3$. In the
renormalization of the parameters $\Delta_1$ and $g_1$ these divergent
terms are absorbed, but one should note that this is the reason for
why the effective range is needed at LO for the P-wave
interaction to be properly renormalized. Matching the dicluster propagator
(\ref{eq:fulldiclusterprop3half1}) to the scattering t-matrix we can
write the propagator using the ERE parameters, scattering length $a_1$
and effective range $r_1$, according to
\begin{equation}
D(E,\mbf{p})=\frac{15\pi}{2mg_1^2}\frac{1}{\frac{1}{a_1}-\frac{1}{2}r_1\left(\frac{8mE}{5}-\frac{4\mbf{p}^2}{25}\right)+i\left(\frac{8mE}{5}-\frac{4\mbf{p}^2}{25}\right)^{3/2}}~.
\label{eq:fulldiclusterprop3half2}
\end{equation}
The $3/2^-$ P-wave scattering diagram is given in
Fig.~\ref{fig:twobodyprops}(b) where we also define the $3/2^-$
dicluster propagator as the double line with internal bottom-to-top
right-tilted lines.

The $1/2^-$ dicluster propagator is given in the same fashion as
\begin{equation}
\tilde{D}(E,\mbf{p})=\frac{15\pi}{2m\tilde{g}_1^2}\frac{1}{\frac{1}{\tilde{a}_1}-\frac{1}{2}\tilde{r}_1\left(\frac{8mE}{5}-\frac{4\mbf{p}^2}{25}\right)+i\left(\frac{8mE}{5}-\frac{4\mbf{p}^2}{25}\right)^{3/2}}~,
\label{eq:fulldiclusterprop1half2}
\end{equation}
with $\tilde{a}_1$ and $\tilde{r}_1$ the corresponding scattering
length and effective range, respectively.
The $1/2^-$ scattering diagram is shown in
Fig.~\ref{fig:twobodyprops}(c), where the double line with internal
bottom-to-top left-tilted lines defines the $1/2^-$ dicluster
propagator.

We extract the ERE parameters for the P-wave channels by matching the
$\nuc{5}{He}$ resonances to the pole positions of the dicluster
propagators (\ref{eq:fulldiclusterprop3half2}) and
(\ref{eq:fulldiclusterprop1half2}). The resulting values are
$a_1=-76.12~\mrm{fm}^3$, $r_1=-141.84~\mrm{MeV}$,
$\tilde{a}_1=-60.37~\mrm{fm}^3$ and
$\tilde{r}_1=66.87~\mrm{MeV}$. Note however that the denominators of
the dicluster propagators are given by third order polynomials. As
such, there is also an unphysical pole for both the $3/2^-$ and
$1/2^-$ channel. These poles need to be removed if we are to perform
three-body calculations using such dicluster propagators.

We now turn to the discussion of three different prescriptions of how
to handle the unphysical pole of a P-wave propagator. The first method
we denote as the unitarity piece removal prescription (UP), since this
presciption removes the $ik^3$ term in the denominator of the
dicluster propagator. The reason for why this is permissable is that
at low momentum the unitarity piece scales as $ik^3\sim\klo^3$ while
the other two terms scale according to
$1/a_1\sim\klo^2\khi\sim r_1k^2$, by assumption. Therefore, at LO, we
may neglect the unitarity piece. Note however that this prescription
is only valid if the scalings of the ERE parameters are as
stated. Otherwise the unphysical pole must be handled in some other
way. The resulting propagator in the UP is
\begin{equation}
D^{(\mrm{UP})}(E,\mbf{p})=\frac{15\pi}{2mg_1^2}\frac{1}{\frac{1}{a_1}-\frac{1}{2}r_1k^2}~,
\label{eq:fulldiclusterprop3half2UP}
\end{equation}
where $k=\sqrt{\frac{8mE}{5}-\frac{4\mbf{p}^2}{25}}$.
As can be seen, the UP has moved the physical pole to the real
momentum axis. This change of the pole position should be within the
bounds of the LO model. Moreover, note that the low-momentum physics
of the UP propagator is unchanged up to order $k^2$.

The second method is the subtraction prescription (SP), where the
unphysical pole is simply subtracted from the dicluster
propagator. The SP propagator is given by
\begin{equation}
D^{(\mrm{SP})}(E,\mbf{p})=\frac{15\pi}{2mg_1^2}\frac{1}{\frac{1}{a_1}-\frac{1}{2}r_1k^2+ik^3}-\frac{\mathcal{R}_0}{k-k_0}~,
\label{eq:fulldiclusterprop3half2SP}
\end{equation}
where $k_0$ and
$\mathcal{R}_0$ are the pole position and residue of the unphysical
pole. There are two things to note about the SP: First, the
prescription does not move the physical pole and, second, it changes
the low-momentum physics. This second point means that the scattering
length and effective range of the SP propagator
(\ref{eq:fulldiclusterprop3half2SP}) are not the same as before the
subtraction. The impact of the subtraction depends on how {\it deep}
the spurious bound state is but can typically considered to be of
higher order.

The third prescription we call the expansion prescription (EP). In the
EP we expand the unphysical pole to second order in momentum and
therefore it does neither change the low-momentum physics, nor move
the physical pole position. The EP propagator is given by
\begin{align}
D^{(\mrm{EP})}(E,\mbf{p})=&\frac{15\pi}{2mg_1^2}\frac{1}{\frac{1}{a_1}-\frac{1}{2}r_1k^2+ik^3}
\nonumber\\
&-\mathcal{R}_0\left(\frac{1}{k-k_0}+\frac{1}{k_0}+\frac{k}{k_0^2}+\frac{k^2}{k_0^3}\right)~.
\label{eq:fulldiclusterprop3half2EP}
\end{align}
As can be seen in Eq.~(\ref{eq:fulldiclusterprop3half2EP}) this
propagator has very different large-momentum asymptotics than the
origanal dicluster propagator, since it now scales as $k^2$ for large
$k$. From an EFT perspective this is not an issue, since the
large-momentum asymptotics is renormalized. However, numerically we
are limited to lower cutoffs in the EP, than in the UP or SP, since
the three-body amplitudes will consist of differences of very large
numbers when the cutoff is large.

\subsection{Three-body scattering diagrams}

\begin{figure}[t!]
\begin{centering}

\includegraphics*[scale=0.3,angle=0,clip=true]{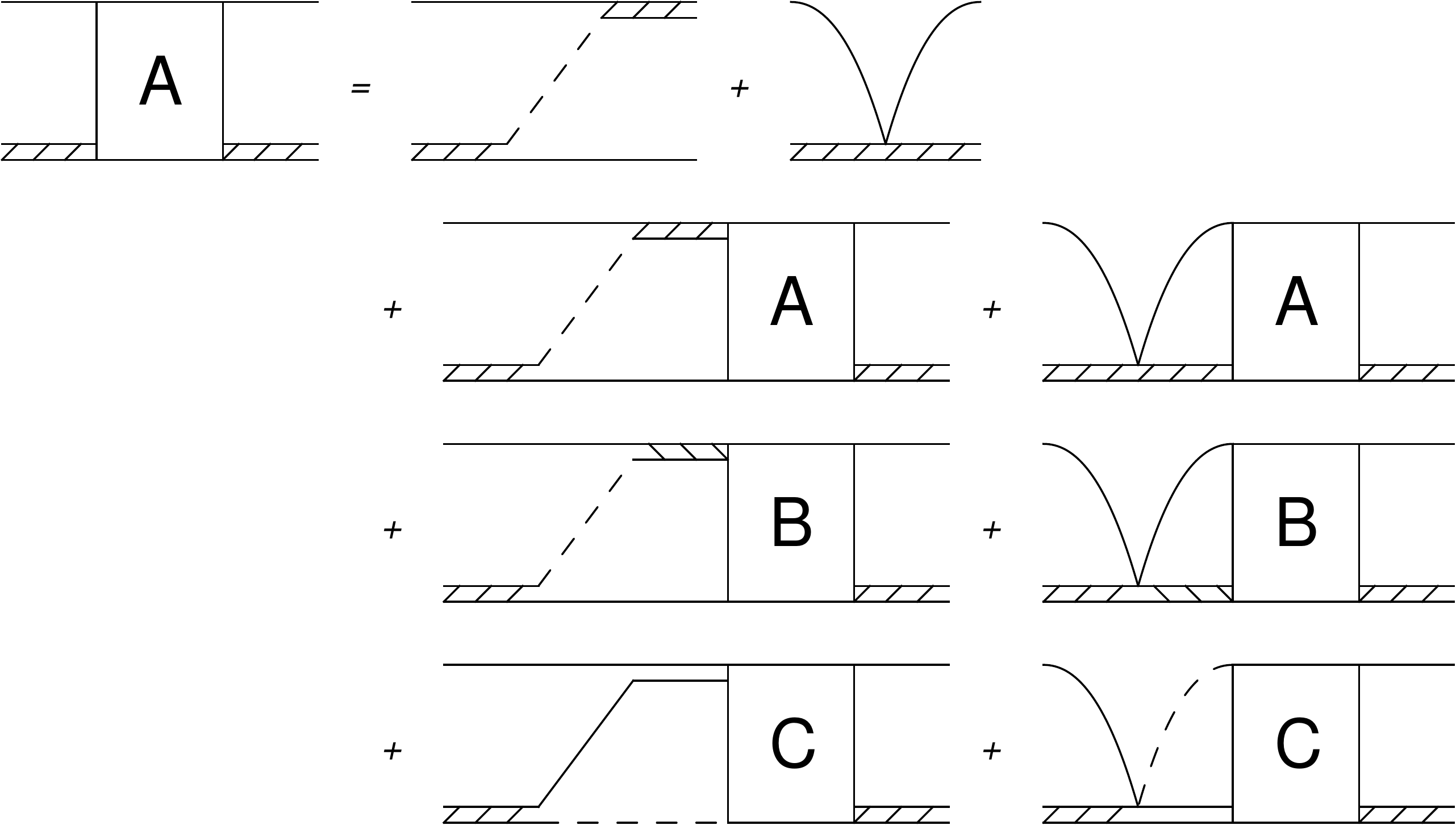}

~

\includegraphics*[scale=0.3,angle=0,clip=true]{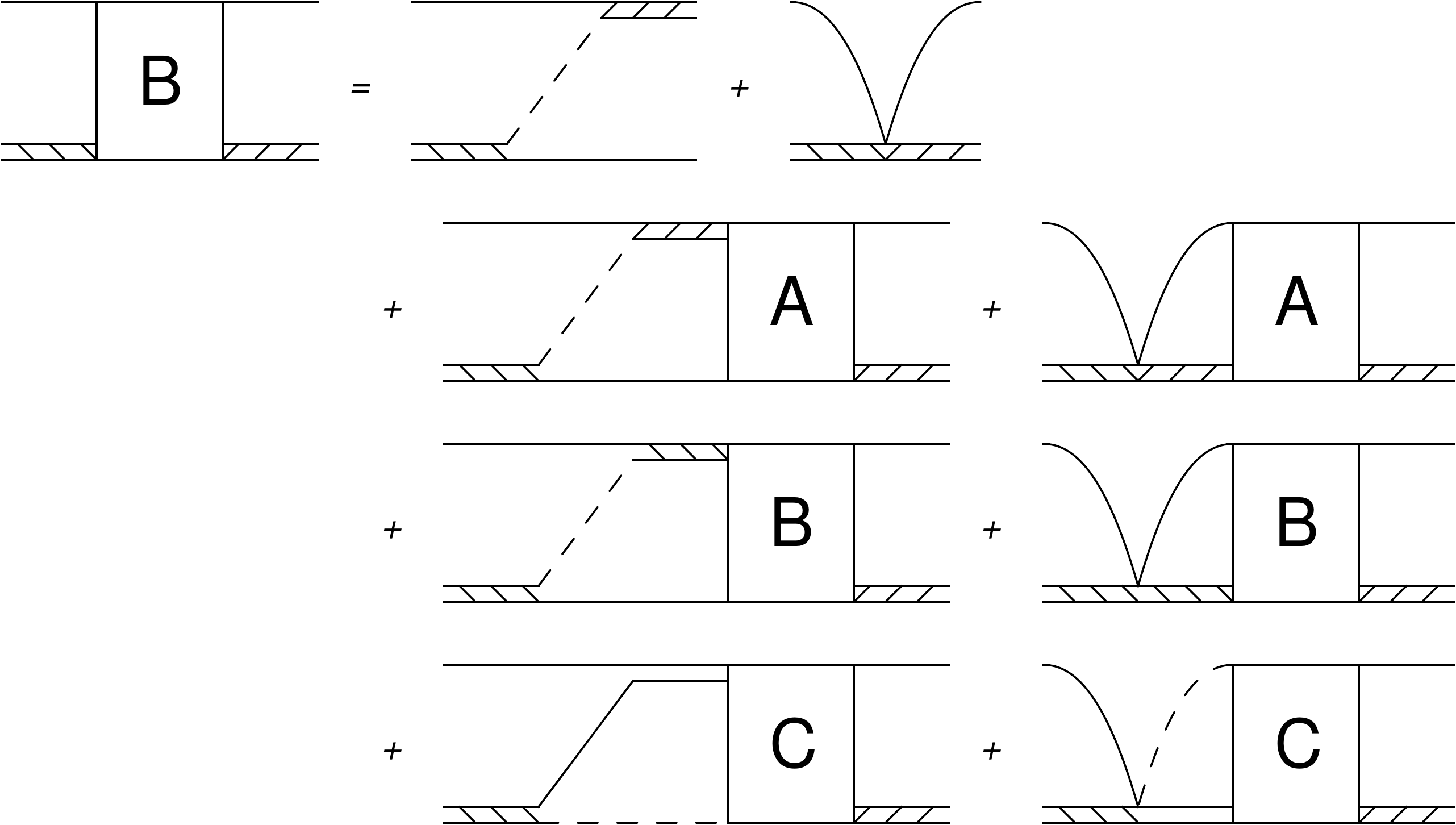}

~

\includegraphics*[scale=0.3,angle=0,clip=true]{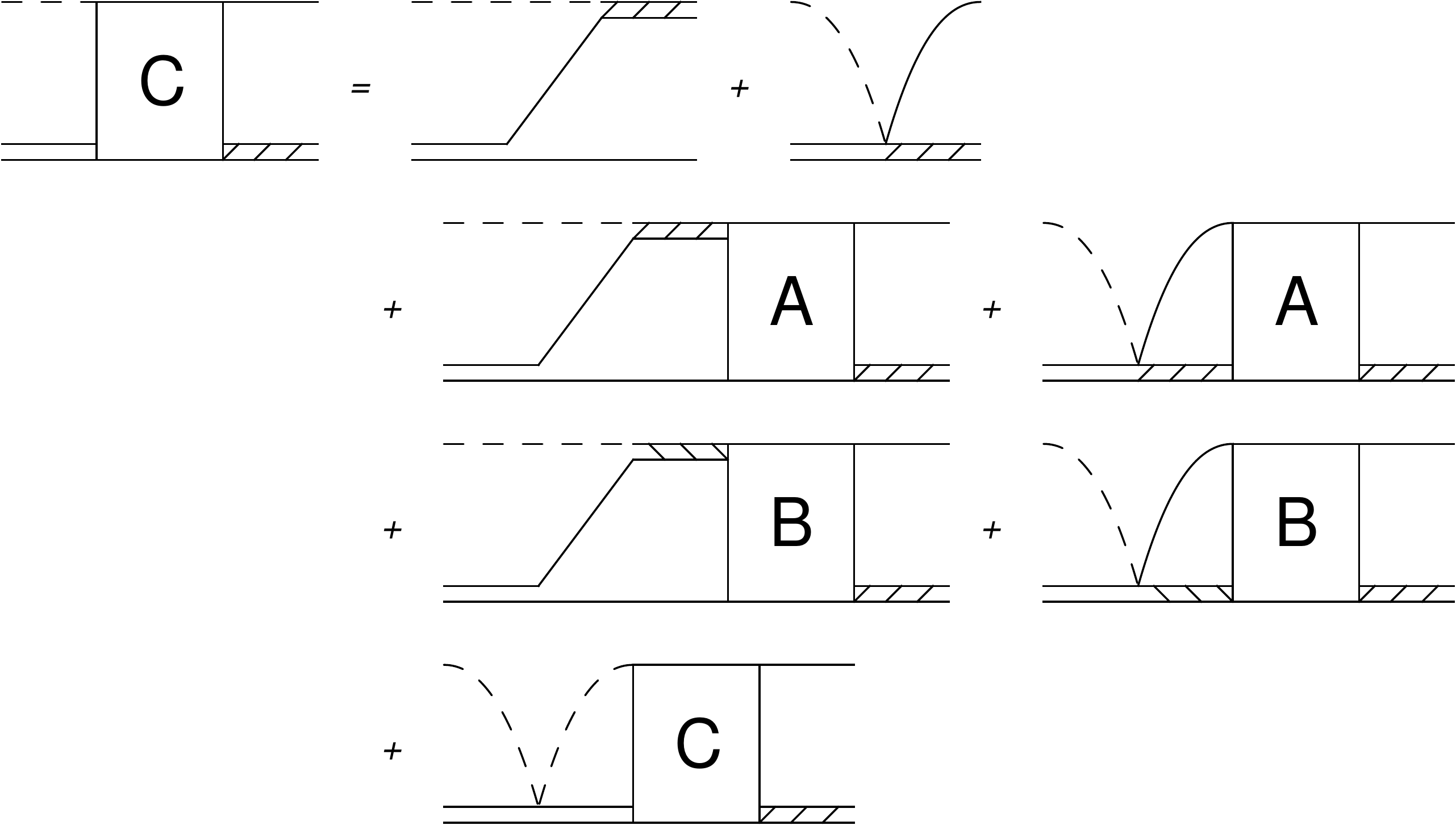}
\caption{The three-body scattering diagrams. Since there are three
  two-body channels the scattering equation is given by a coupled
  integral equation of three parts. See text for details.}
\label{fig:diagram3bodyscattering}
\end{centering}
\end{figure}

In this part we write down the three-body integral equations for the
$0^+$ channel of $\nuc{6}{He}$. We define the integral equations to
have outgoing $\nuc{5}{He}(3/2^-)$ and neutron legs. Since we have
three two-body channels the integral equation is a coupled system with
three parts: (i) The A-amplitude, with incoming $\nuc{5}{He}(3/2^-)$
and neutron legs, (ii) the B-amplitude, with incoming
$\nuc{5}{He}(1/2^-)$ and neutron legs, and (iii) the C-amplitude, with
incoming dineutron and $\alpha$-particle legs. Note that to in order
to have a total $0^+$ the $\nuc{5}{He}$+neutron legs must be in a
relative P-wave, while the dineutron+$\alpha$ legs must be in a
relative S-wave. The three-body integral equations are shown in a
diagramatic form in Fig.~\ref{fig:diagram3bodyscattering}.

On the first line for each of the amplitudes in
Fig.~\ref{fig:diagram3bodyscattering} the two inhomogeneous terms are
shown. However, in this paper we are only concerned with bound-state
solutions for which the scattering
amplitude has a pole for negative energies. As such the inhomogeneous terms are not
necessary for our purpose. We have constructed the
integral equations from the diagrams in
Fig.~\ref{fig:diagram3bodyscattering}. The homogeneous part of the integral equations
projected onto a total $0^+$ is given by
\begin{align}
iA^{(A)}(k,p)=&ig_1^2 \frac{2m}{3\pi^2}\int\d q q^2 K^{(AA)}(k,q) D\left(E-\frac{3q^2}{5m},0\right)A(q,p)
\label{eq:diagram0+AA}
\\
iA^{(B)}(k,p)=&ig_1\tilde{g}_1 \frac{2m}{3\pi^2}\int\d q q^2 K^{(AB)}(k,q)\tilde{D}\left(E-\frac{3q^2}{5m},0\right)B(q,p)
\\
iA^{(C)}(k,p)=&ig_1g_0 \frac{m}{\sqrt{6}\pi^2}\int\d q q^2 K^{(AC)}(k,q) B\left(E-\frac{3q^2}{8m},0\right) C(q,p)
\\
iB^{(A)}(k,p)=&ig_1\tilde{g}_1\frac{2m}{3\pi^2}\int\d q q^2 K^{(BA)}(k,q)D\left(E-\frac{3q^2}{5m},0\right)A(q,p)
\\
iB^{(B)}(k,p)=&i\tilde{g}_1^2\frac{2m}{3\pi^2}\int\d q q^2 K^{(BB)}(k,q)\tilde{D}\left(E-\frac{3q^2}{5m},0\right)B(q,p)
\\
iB^{(C)}(k,p)=&i\tilde{g}_1g_0\frac{m}{\sqrt{6}\pi^2}\int\d q q^2 K^{(BC)}(k,q)B\left(E-\frac{3q^2}{8m},0\right)C(q,p)
\\
iC^{(A)}(k,p)=&2ig_0g_1\frac{m}{\sqrt{6}\pi^2}\int\d q q^2 K^{(CA)}(k,q)D\left(E-\frac{3q^2}{5m},0\right)A(q,p)
\\
iC^{(B)}(k,p)=&2ig_0\tilde{g}_1\frac{m}{\sqrt{6}\pi^2}\int\d q q^2 K^{(CB)}(k,q)\tilde{D}\left(E-\frac{3q^2}{5m},0\right)B(q,p)
\\
iC^{(C)}(k,p)=&2ig_0^2\frac{m}{2\pi^2}\int\d q q^2 K^{(CC)}(k,q) B\left(E-\frac{3q^2}{8m}\right)C(q,p)~.
\label{eq:diagram0+CC}
\end{align}
The momenta are defined such that the incoming fields have relative
momentum $\mbf{k}$ and the outgoing fields have relative momentum
$\mbf{p}$. The loop-momentum $q$ is limited by the cutoff $\Lambda$
and the integration span $[0,\Lambda]$ is replaced with a Legendre
mesh when the integral equations are solved numerically.

The kernels $K^{XY}$, where $X,Y=A,B,C$, are defined as
\begin{align}
K^{(AA)}(k,p)=&\frac{27}{25}Q_0(\rho_1(k,p))+\frac{2}{5}\frac{k^2+p^2}{kp}Q_1(\rho_1(k,p))+Q_2(\rho_1(k,p))
\nonumber\\
&-H_1 \frac{kp}{\Lambda^2}
\label{eq:KAA}\\
K^{(AB)}(k,p)=&K^{(BA)}(k,p)
\nonumber\\
=&\frac{\sqrt{2}}{25}Q_0(\rho_1(k,p))+\frac{\sqrt{2}}{5}\frac{k^2+p^2}{kp}Q_1(\rho_1(k,p))+\sqrt{2}Q_2(\rho_1(k,p))
\nonumber\\
&-H_2\frac{kp}{\Lambda^2}\\
K^{(AC)}(k,p)=&K^{(CA)}(p,k)
\nonumber\\
=&\frac{4}{5p}Q_0(\rho_2(k,p))+\frac{1}{k}Q_1(\rho_2(k,p))-H_5\frac{k}{\Lambda^2}
\\
K^{(BB)}(k,p)=&\frac{26}{25}Q_0(\rho_1(k,p))+\frac{1}{5}\frac{k^2+q^2}{kq}Q_1(\rho_1(k,q))-H_3\frac{kp}{\Lambda^2}
\\
K^{(BC)}(k,p)=&K^{(CB)}(p,k)
\nonumber\\
=&\frac{4}{5\sqrt{2}p}Q_0(\rho_2(k,p))+\frac{1}{\sqrt{2}k}Q_1(\rho_2(k,p))-H_6\frac{k}{\Lambda^2}
\\
K^{(CC)}(k,p)=&-H_4\frac{1}{\Lambda^2}~,
\label{eq:KCC}
\end{align}
where the arguments to the Legendre-Q functions $Q_L$ are given by
\begin{align}
\rho_1(k,p)=&\frac{4mE}{kp}-\frac{5k}{2p}-\frac{5p}{2k}\\
\rho_2(k,p)=&\frac{mE}{kp}-\frac{k}{p}-\frac{5p}{8k}~.
\end{align}
For convenience we have introduced new three-body parameters and these
are given in terms of the old ones as
\begin{align}
H_1=&\frac{h_1}{4m g_1^2}\Lambda^2\\
H_2=&\frac{h_2}{4m g_1 \tilde{g}_1}\Lambda^2\\
H_3=&\frac{h_3}{4m \tilde{g}_1^2}\Lambda^2\\
H_4=&\frac{h_4}{m g_0^2}\Lambda^2\\
H_5=&\frac{h_5}{\sqrt{2}m g_0g_1}\Lambda^2\\
H_6=&\frac{h_6}{mg_0\tilde{g}_1}\Lambda^2~.
\end{align}
Note that the $H_1$, $H_3$ and $H_4$ are what we refer to as the
diagonal three-body interactions, while $H_2$, $H_5$ and $H_6$ are the
off-diagonal ones.

\section{Renormalization of bound states}
\label{sec:results}
%

In this section we present results for the renormalization of the
$\nuc{6}{He}(0^+)$ bound state, using the six three-body interactions
at the lowest scaling dimension and the three different prescriptions
for how to handle the unphysical two-body poles.

In searching for the bound state we search for negative-energy solutions to the
eigenvalue matrix equation V=KV, where V is the vector of amplitudes
and K is the kernel matrix. The kernel matrix is contructed by
numerically replacing the momenta by Legendre meshes over
$[0,\Lambda]$. The solution is obtained by finding the zero of the
kernel determinant, that is
\begin{equation}
\det\left(\mbf{1}-K\right)=0~.
\label{eq:kerneldet}
\end{equation}

\subsection{Field theory with dineutron and $\nuc{5}{He}(3/2^-)$ channels}

\begin{figure}[t!]
\begin{centering}
\includegraphics*[width=0.6\textwidth,angle=0,clip=true]{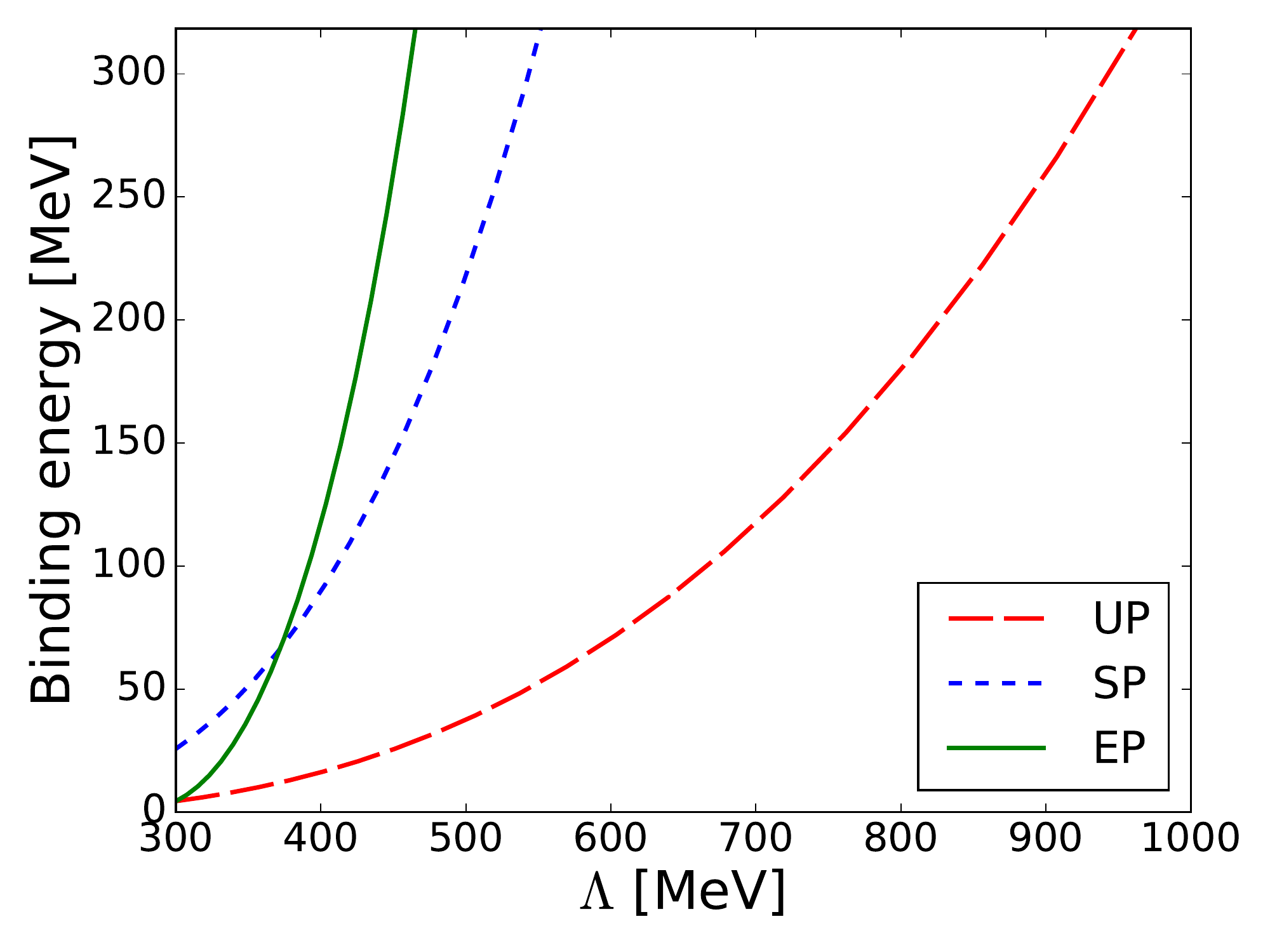}
\caption{The $\nuc{6}{He}$ binding energy ($B=-E$) as a function of the cutoff
  $\Lambda$. Only the dineutron and the $\nuc{5}{He}(3/2^-)$ channels are
  included. All the three-body interactions have been set to zero.}
\label{fig:BLambda_only3half}
\end{centering}
\end{figure}

\begin{figure}[t!]
\includegraphics*[width=\textwidth,angle=0,clip=true]{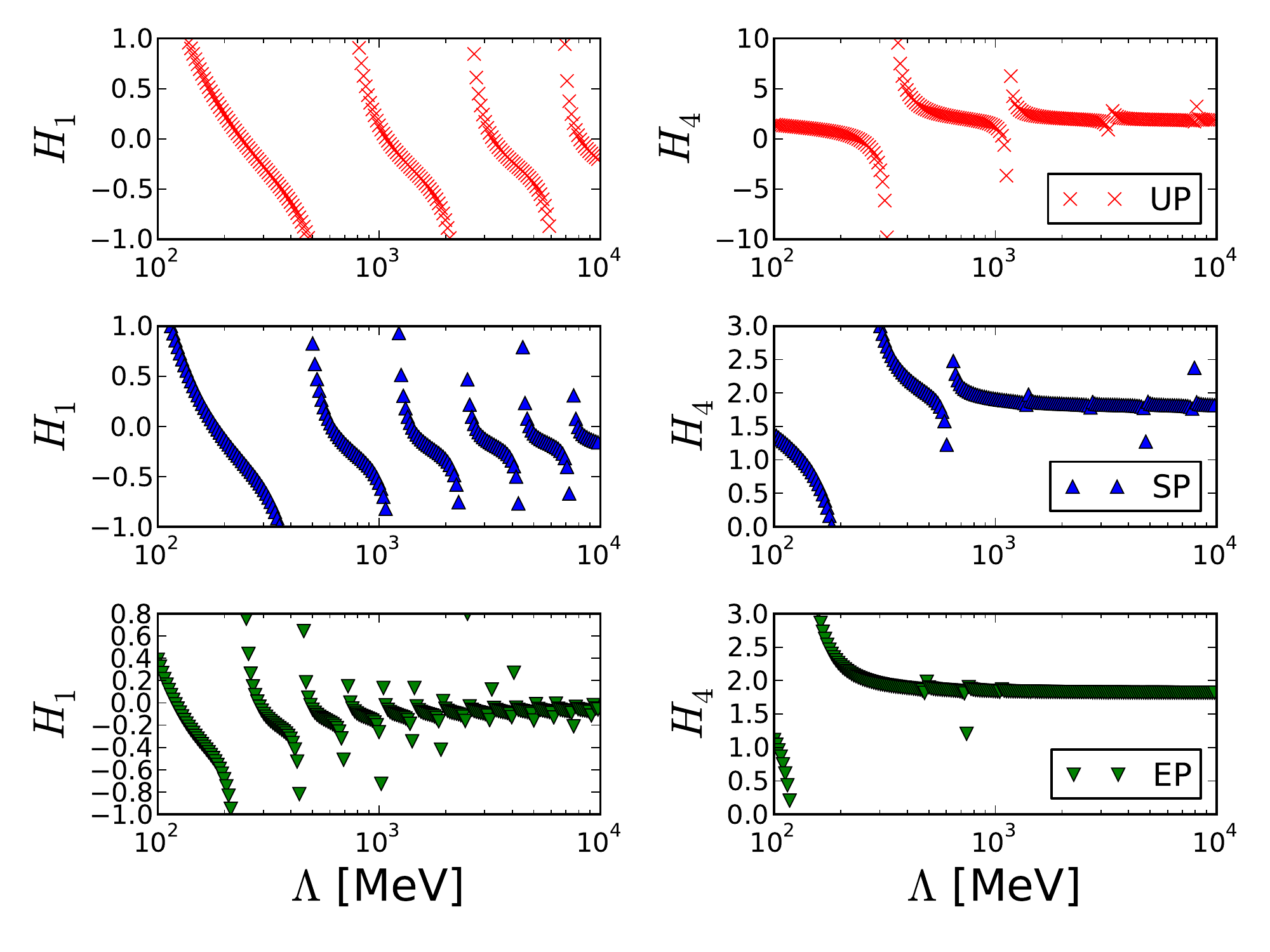}
\caption{The running of the three-body forces $H_1$ (left panels) and
  $H_4$ (right panels) for the three different pole removal
  prescriptions. Only the dineutron and the $\nuc{5}{He}(3/2^-)$
  channels are included. The three-body parameter was fixed to
  reproduce the bound-state energy $E=-1~\mrm{MeV}$.}
\label{fig:HLambda_only3half}
\end{figure}

\begin{figure}[t!]
\includegraphics*[width=\textwidth,angle=0,clip=true]{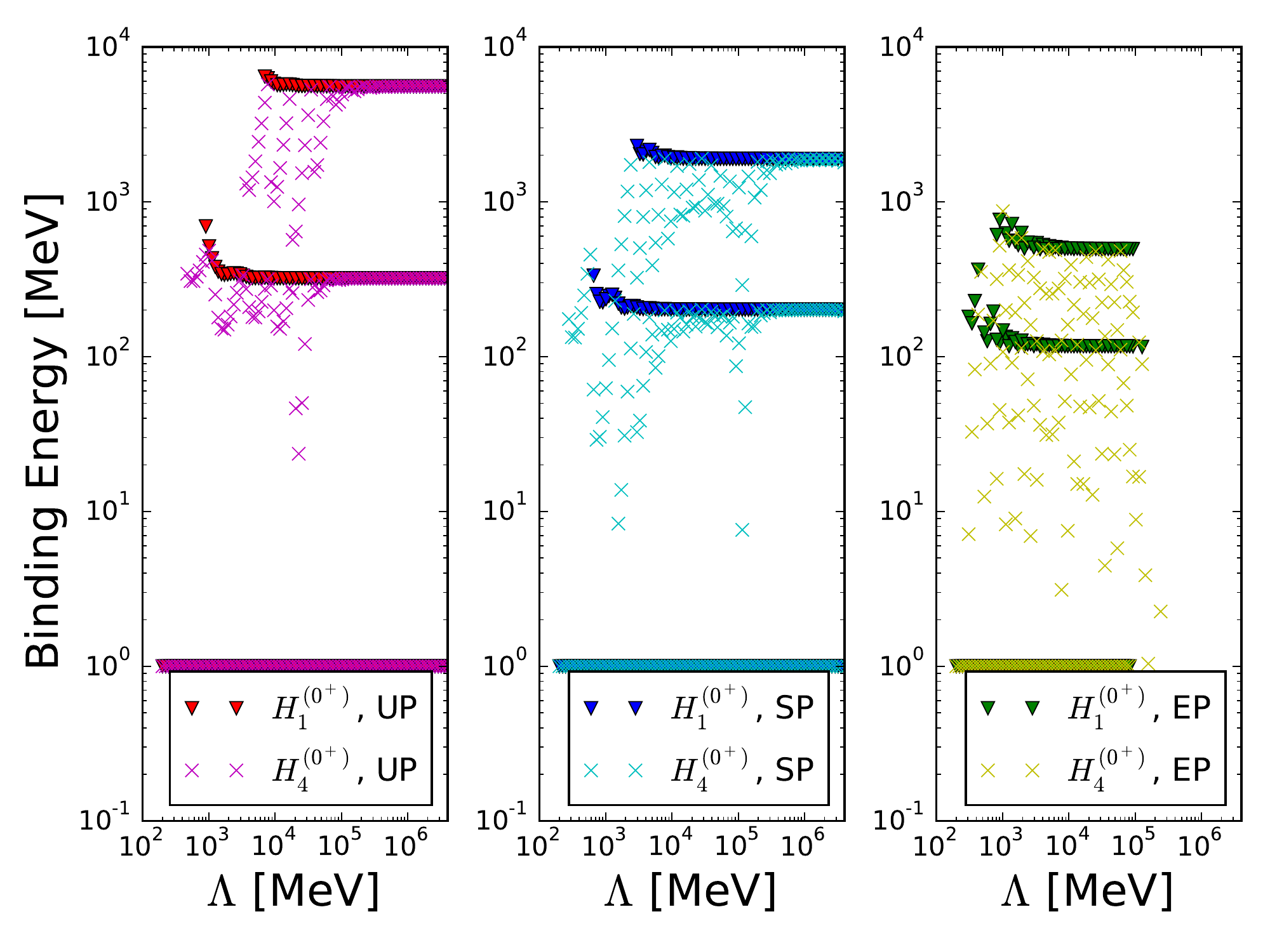}
\caption{Convergence of deep bound states for the three different
  prescriptions, using either the $H_{1}$ or $H_{4}$ three-body
  interaction. Only the dineutron and the $\nuc{5}{He}(3/2^-)$ channels are
  included. The three-body parameter was fixed to reproduce the bound-state
  energy $E=-1~\mrm{MeV}$.}
\label{fig:manyboundstates_only3half}
\end{figure}

We begin by using the simpler field theory, where only the dineutron
and the $\nuc{5}{He}(3/2^-)$ channels are included. First, we set all
the three-body interactions for this field theory, $H_1$, $H_4$ and
$H_5$, to zero and evaluate the integral equations for different
cutoffs. We performed these calculations using all three prescriptions
(UP, SP, and EP) for the handling of the unphysical two-body
poles. From this we obtain the cutoff dependence of the bound-state
energy. The result of this numerical study is presented in
Fig.~\ref{fig:BLambda_only3half}. As can be seen, the bound-state
energy is cutoff dependent and this implies that (at least) one
three-body interaction is needed to renormalize the field theory.

We then move on to solving the integral equations for a fixed binding
energy, $E=-1~\mrm{MeV}$, and one varying three-body
interaction. Thus, for each value of the cutoff we find the value of a
selected three-body parameter that gives the fixed binding energy. For
the off-diagonal three-body interaction, $H_5$, we did not find any
solutions to the integral equation. This indicates that it can not
renormalize the field theory. Using the diagonal interactions, $H_1$
and $H_4$, we generated the results, shown in
Fig.~\ref{fig:HLambda_only3half}. As before, we performed these
calculations for all three unphysical-pole-removal prescriptions. It
is seen that the renormalization shows a limit-cycle like
behavior for both the $H_1$ and $H_4$ three-body interactions. The
poles that can be seen in Fig.~\ref{fig:HLambda_only3half} are
associated with the appearance of additional three-body bound
states. The different pole removal prescriptions differ in the amount
of states that are generated at larger energies since the different
pole removal prescriptions lead to different large momentum behaviors
in the two-body amplitude.

Finally, we also use the dineutron and $\nuc{5}{He}(3/2^-)$ field
theory to search for deep bound states in the $0^+$ channel. Of
course, these states are not true states of $\nuc{6}{He}$, but they
are still well-defined observables in the field theory. As such, if
these deep bound states do not converge with increasing cutoff, then
that indicates that the field theory is not properly renormalized. The
procedure is as follows: First we fix either the $H_1$ or $H_4$
three-body interaction to reproduce the bound-state energy
$E=-1~\mrm{MeV}$ for a fixed value of the cutoff. Then, we use this
three-body interaction to search for additional solutions to
Eq.~\eqref{eq:kerneldet} for larger binding energies. These solutions
are then deep bound states of the system. Third, we repeat the process
for larger values of the cutoff. The result is then a convergence plot
of these deep bound states and it is given in
Fig.~\ref{fig:manyboundstates_only3half}. Five findings are of
particular note regarding the convergence of these deep-bound states:
(i) The deep-bound states do indeed converge for large cutoffs, (ii) the
convergence using the $H_1$ three-body interaction is much faster than
using the $H_4$ one, (iii) the two different three-body interactions
renormalize the deep bound states to the same binding energy, (iv) the
three different prescriptions produce a different bound-state
spectrum, and (v) using the EP we are numerically limited to cutoffs
$\Lambda\lesssim10^5~\mrm{MeV}$ and for these cutoffs the deep bound
states have not yet converged when the $H_4$ interaction is
used. Summarizing this part, it is very encouraging that the $H_1$ and
$H_4$ interactions give the same spectrum and since the $H_1$ produces
faster convergence this motivates the use of only this three-body
interaction. This conclusion is in line with previous work on the
$\nuc{6}{He}(0^+)$, where only this three-body interaction was
considered, see Refs.~\cite{Rotureau:2012yu,Ji:2014wta}.

\subsection{Field theory with dineutron, $\nuc{5}{He}(3/2^-)$ and $\nuc{5}{He}(1/2^-)$ channels}

\begin{figure}[t!]
\begin{centering}
\includegraphics*[width=\textwidth,angle=0,clip=true]{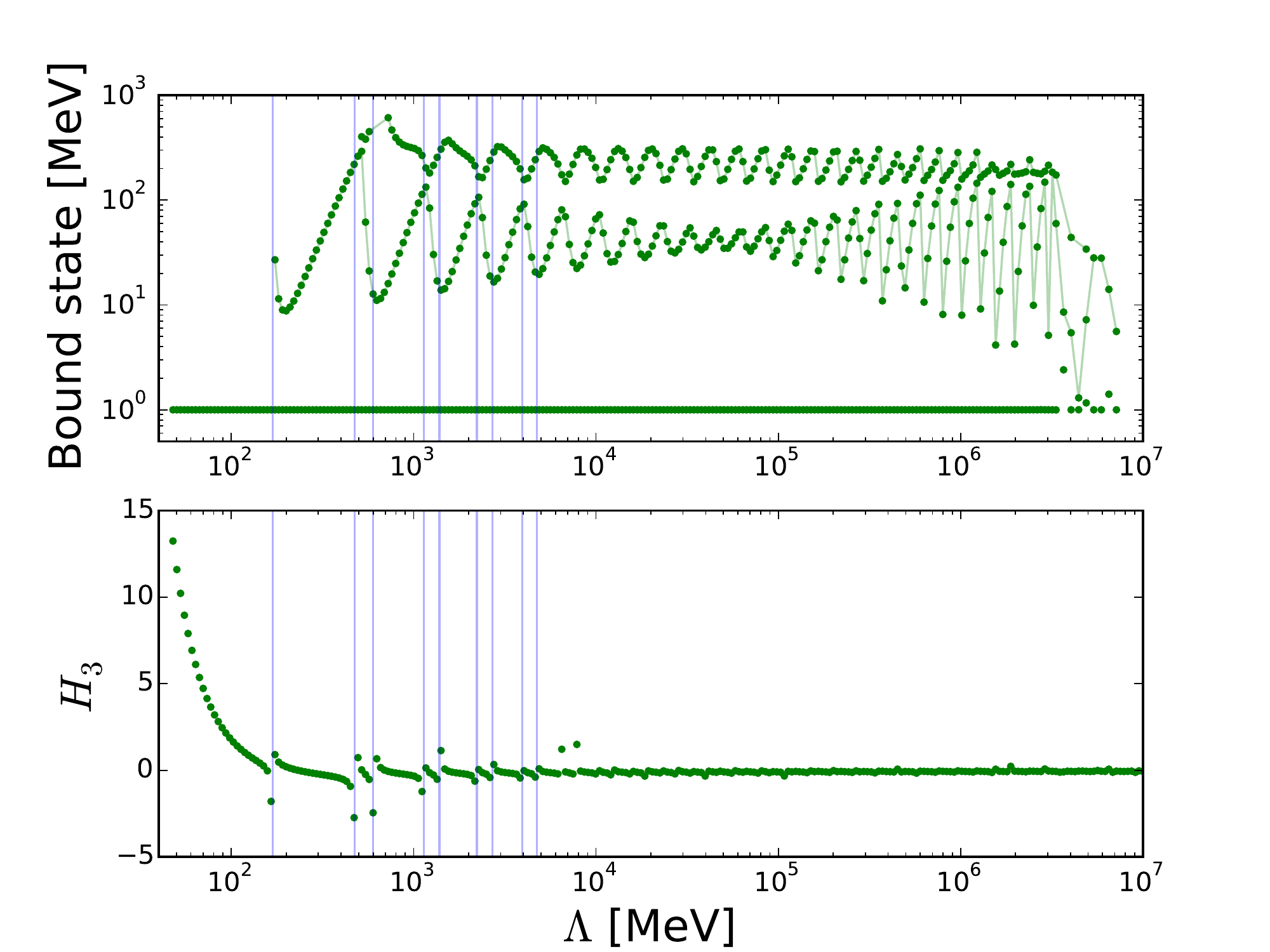}
\caption{Non-convergence of deep bound states for a field theory where
  the $1/2^-$ channel is also included. 
Only the $H_3$ three-body interaction is used.
The three-body parameter was fixed to the bound state energy $B=1~\mrm{MeV}$. The vertical blue lines indicate singularities in the limit cycle and it can be seen that the oscillation of the deep bound states are given by the period of these singularities.}
\label{fig:manyboundstates_3halfand1half}
\end{centering}
\end{figure}

We now include also the $\nuc{5}{He}(1/2^-)$ channel into the field
theory. As such we have three channels to consider, which
increases the computational cost by about a factor of two compared to
the two-channel system. Further, there are now in total six three-body
interactions at the lowest scaling dimension. However, in solving the
integral equations for a bound state we did not find any solutions for
the off-diagonal three-body interactions $H_2$, $H_5$ and $H_6$.

When a search for deep bound states was performed, using one of the
diagonal three-body interactions $H_1$, $H_3$ and $H_4$ at a time, no
convergence was observed. The cutoff dependence of the deep-bound
states, using the SP and the $H_3$ three-body interaction, can be seen in
Fig.~\ref{fig:manyboundstates_3halfand1half}. It is clear that these
deep bound states are much more shallow than in the field theory
without the $\nuc{5}{He}(1/2^-)$ and we are limited to a lower cutoff,
$\Lambda\lesssim10^6~\mrm{MeV}$. This indicates that the integral
equation for the field theory with the $\nuc{5}{He}(1/2^-)$ included
is more involved numerically. The result implies that one three-body
interaction is not sufficient for proper renormalization of the field
theory. As such one would need to fix two, or more, three-body interactions
simultaneously. However, it could also be the case that the deep bound
states are renormalized at a larger cutoff than is presently accessible to us.

%
\section{Conclusion}
\label{sec:conclusion}

In this paper we have derived the LO integral equations for the
treatment of the $0^+$ channel of $\nuc{6}{He}$ from quantum field
theory. We included not only the dineutron and $\nuc{5}{He}(3/2^-)$
channels, but also the $\nuc{5}{He}(1/2^-)$. We analyzed all six
three-body interactions that appear at the lowest mass dimension.
Further, we discussed three different
prescriptions of how to handle the unphysical pole in the P-wave
$\nuc{5}{He}$ dicluster propagators. [[We showed that all three methods
are useful but that they lead to different convergence behavior when
employed in the few-body sector.]]

For a field theory where only the dineutron and the
$\nuc{5}{He}(3/2^-)$ channels were included we showed that the system
was properly renormalized. In addition, both the $H_1$ and $H_4$
three-body interactions generated the same bound state
spectrum. However, for a field theory where the $\nuc{5}{He}(1/2^-)$
was also included the system was not renormalizable, at least not when
only one three-body interaction was used and for cutoffs
$\Lambda\lesssim10^6~\mrm{MeV}$.

Future studies will concern the resonant spectrum and additional total
angular momentum channels, for example the $\nuc{6}{He}(2^+)$.
Our results are also relevant for future applications of Halo EFT to
D-wave systems such as low-lying resonances in the Oxygen-25 and
Oxygen-26 nuclei. In these systems, Oxygen-24 is interacting strongly
in a relative D-wave with the neutrons. Furthermore, additional
spurious poles are expected in the two-body sector due to the
polynomial structure of the effective range expansion that appears
after renormalization in the denominator of two-body propagators.

%
\begin{acknowledgements}
  This work was supported by the Swedish Research Council
  (dnr. 2010-4078), the European Research Council under the European
  Community's Seventh Framework Programme (FP7/2007-2013) / ERC grant
  agreement no.~240603, the Swedish Foundation for International
  Cooperation in Research and Higher Education (STINT, IG2012-5158),
  the National Science Foundation under Grant No.  PHY-1555030, and by
  the Office of Nuclear Physics, U.S.~Department of Energy under under
  Contract No.\ DE-AC05-00OR2272.
\end{acknowledgements}

\bibliographystyle{myspbasic}
\bibliography{3b_refs}

\begin{thebibliography}{26}
\providecommand{\natexlab}[1]{#1}
\providecommand{\url}[1]{{#1}}
\providecommand{\urlprefix}{URL }
\expandafter\ifx\csname urlstyle\endcsname\relax
  \providecommand{\doi}[1]{DOI~\discretionary{}{}{}#1}\else
  \providecommand{\doi}{DOI~\discretionary{}{}{}\begingroup
  \urlstyle{rm}\Url}\fi
\providecommand{\eprint}[2][]{\url{#2}}

\bibitem[{Epelbaum et~al(2009)Epelbaum, Hammer, and Meissner}]{Epelbaum:2008ga}
Epelbaum E, Hammer HW, Meissner UG (2009) {Modern Theory of Nuclear Forces}.
  Rev Mod Phys 81:1773--1825, \doi{10.1103/RevModPhys.81.1773},
  \eprint{0811.1338}

\bibitem[{Machleidt and Entem(2011)}]{Machleidt:2011zz}
Machleidt R, Entem DR (2011) {Chiral effective field theory and nuclear
  forces}. Phys Rept 503:1--75, \doi{10.1016/j.physrep.2011.02.001},
  \eprint{1105.2919}

\bibitem[{Braaten and Hammer(2006)}]{Braaten:2004rn}
Braaten E, Hammer HW (2006) {Universality in few-body systems with large
  scattering length}. Phys Rept 428:259--390,
  \doi{10.1016/j.physrep.2006.03.001}, \eprint{cond-mat/0410417}

\bibitem[{Hammer et~al(2013)Hammer, Nogga, and Schwenk}]{Hammer:2012id}
Hammer HW, Nogga A, Schwenk A (2013) {Three-body forces: From cold atoms to
  nuclei}. Rev Mod Phys 85:197, \doi{10.1103/RevModPhys.85.197},
  \eprint{1210.4273}

\bibitem[{Tanihata et~al(1985)Tanihata, Hamagaki, Hashimoto, Shida, Yoshikawa
  et~al}]{Tanihata:1986kh}
Tanihata I, Hamagaki H, Hashimoto O, Shida Y, Yoshikawa N, et~al (1985)
  {Measurements of Interaction Cross-Sections and Nuclear Radii in the Light p
  Shell Region}. Phys Rev Lett 55:2676--2679

\bibitem[{Jonson(2004)}]{2004PhR...389....1J}
Jonson B (2004) {Light dripline nuclei}. Phys Rep 389:1--59

\bibitem[{Bertulani et~al(2002)Bertulani, Hammer, and van
  Kolck}]{Bertulani:2002sz}
Bertulani C, Hammer HW, van Kolck U (2002) {Effective field theory for halo
  nuclei}. Nucl Phys A 712:37--58

\bibitem[{Bedaque et~al(2003)Bedaque, Hammer, and van Kolck}]{Bedaque:2003wa}
Bedaque PF, Hammer HW, van Kolck U (2003) {Narrow resonances in effective field
  theory}. Phys Lett B569:159--167

\bibitem[{Higa et~al(2008)Higa, Hammer, and van Kolck}]{Higa:2008dn}
Higa R, Hammer HW, van Kolck U (2008) {$\alpha$$\alpha$ scattering in halo
  effective field theory}. Nucl Phys A 809:171

\bibitem[{Brown and Hale(2014)}]{Brown:2013zla}
Brown LS, Hale GM (2014) {Field Theory of the d+t$\to$n+alpha Reaction
  Dominated by a 5He* Unstable Particle}. Phys Rev C89:014,622,
  \eprint{1308.0347}

\bibitem[{Hammer and Phillips(2011)}]{Hammer:2011ye}
Hammer HW, Phillips DR (2011) {Electric properties of the Beryllium-11 system
  in Halo EFT}. Nucl Phys A 865:17--42

\bibitem[{Rupak et~al(2012)Rupak, Fernando, and Vaghani}]{Rupak:2012cr}
Rupak G, Fernando L, Vaghani A (2012) {Radiative Neutron Capture on Carbon-14
  in Effective Field Theory}. Phys Rev C86:044,608, \eprint{1204.4408}

\bibitem[{Acharya and Phillips(2013)}]{Acharya:2013nia}
Acharya B, Phillips DR (2013) {Carbon-19 in Halo EFT: Effective-range
  parameters from Coulomb-dissociation experiments}. Nucl Phys A913:103--115

\bibitem[{Zhang et~al(2014)Zhang, Nollett, and Phillips}]{Zhang:2013kja}
Zhang X, Nollett KM, Phillips DR (2014) {Marrying ab initio calculations and
  Halo-EFT: the case of ${}^7{\rm Li} + n \rightarrow {}^8{\rm Li} + \gamma$}.
  Phys Rev C89(2):024,613, \doi{10.1103/PhysRevC.89.024613}, \eprint{1311.6822}

\bibitem[{Ryberg et~al(2014)Ryberg, Forss\'en, Hammer, and Platter}]{paper:A}
Ryberg E, Forss\'en C, Hammer HW, Platter L (2014) {Effective field theory for
  proton halo nuclei}. Phys Rev C89(1):014,325,
  \doi{10.1103/PhysRevC.89.014325}, \eprint{1308.5975}

\bibitem[{Zhang et~al(2014)Zhang, Nollett, and Phillips}]{Zhang:2014zsa}
Zhang X, Nollett KM, Phillips DR (2014) {Combining ab initio calculations and
  low-energy effective field theory for halo nuclear systems: The case of
  ${}^7Be+p \to {}^8B+\gamma$}. Phys Rev C89(5):051,602,
  \doi{10.1103/PhysRevC.89.051602}, \eprint{1401.4482}

\bibitem[{Ryberg et~al(2014)Ryberg, Forss\'en, Hammer, and Platter}]{paper:B}
Ryberg E, Forss\'en C, Hammer HW, Platter L (2014) {Constraining Low-Energy
  Proton Capture on Beryllium-7 through Charge Radius Measurements}. Eur Phys J
  A50:170, \doi{10.1140/epja/i2014-14170-2}, \eprint{1406.6908}

\bibitem[{Ryberg et~al(2016)Ryberg, Forss\'en, Hammer, and Platter}]{paper:C}
Ryberg E, Forss\'en C, Hammer HW, Platter L (2016) {Range corrections in Proton
  Halo Nuclei}. Annals Phys 367:13--32, \doi{10.1016/j.aop.2016.01.008},
  \eprint{1507.08675}

\bibitem[{Rotureau and van Kolck(2013)}]{Rotureau:2012yu}
Rotureau J, van Kolck U (2013) {Effective Field Theory and the Gamow Shell
  Model: The $\nuc{6}{He}$ Halo Nucleus}. Few Body Syst 54:725--735

\bibitem[{Ji et~al(2014)Ji, Elster, and Phillips}]{Ji:2014wta}
Ji C, Elster C, Phillips DR (2014) {$^6$He nucleus in halo effective field
  theory}. Phys Rev C90(4):044,004, \doi{10.1103/PhysRevC.90.044004},
  \eprint{1405.2394}

\bibitem[{Acharya et~al(2013)Acharya, Ji, and Phillips}]{Acharya:2013aea}
Acharya B, Ji C, Phillips D (2013) {Implications of a matter-radius measurement
  for the structure of Carbon-22}. Phys Lett B723:196--200, \eprint{1303.6720}

\bibitem[{Hagen et~al(2013{\natexlab{a}})Hagen, Hammer, and
  Platter}]{Hagen:2013xga}
Hagen P, Hammer HW, Platter L (2013{\natexlab{a}}) {Charge form factors of
  two-neutron halo nuclei in halo EFT}. Eur Phys J A49:118, \eprint{1304.6516}

\bibitem[{Hagen et~al(2013{\natexlab{b}})Hagen, Hagen, Hammer, and
  Platter}]{Hagen:2013jqa}
Hagen G, Hagen P, Hammer HW, Platter L (2013{\natexlab{b}}) {Efimov Physics
  Around the Neutron-Rich $^{60}$Ca Isotope}. Phys Rev Lett 111(13):132,501,
  \eprint{1306.3661}

\bibitem[{Bedaque et~al(1999)Bedaque, Hammer, and van Kolck}]{Bedaque:1998km}
Bedaque PF, Hammer HW, van Kolck U (1999) {The Three boson system with short
  range interactions}. Nucl Phys A646:444--466,
  \doi{10.1016/S0375-9474(98)00650-2}, \eprint{nucl-th/9811046}

\bibitem[{Tilley et~al(2002)Tilley, Cheves, Godwin, Hale, Hofmann, Kelley,
  Sheu, and Weller}]{Tilley:2002vg}
Tilley DR, Cheves CM, Godwin JL, Hale GM, Hofmann HM, Kelley JH, Sheu CG,
  Weller HR (2002) {Energy levels of light nuclei A=5, A=6, A=7}. Nucl Phys
  A708:3--163, \doi{10.1016/S0375-9474(02)00597-3}

\bibitem[{G{\aa}rdestig(2009)}]{Gardestig:2009ya}
G{\aa}rdestig A (2009) {Extracting the neutron-neutron scattering length -
  recent developments}. J Phys G36:053,001,
  \doi{10.1088/0954-3899/36/5/053001}, \eprint{0904.2787}

\end{thebibliography}

\end{document}